\documentclass[12pt]{article}
\begin{document}

\addtolength{\baselineskip}{0.5\baselineskip}

\title{\textbf{On Emerging Field of Quantum Chemistry at Finite Temperature}}
\author{Liqiang Wei\\
Institute for Theoretical Atomic, Molecular and Optical Physics\\
Harvard University, Cambridge, MA 02138}
 \maketitle

\begin{abstract}
\vspace{0.05in}
  In this article, we present an emerging field of quantum chemistry at finite
 temperature. We discuss its recent developments on both theoretical and experimental fronts.
 We describe and analyze several experimental investigations related
 to the temperature effects on the structure, electronic spectra,
 or bond rupture forces for molecules. This includes the study of the temperature
 impact on the pathway shifts for the protein unfolding by atomic force
 microscopy ($AFM$), the temperature dependence of the absorption spectra of electrons in solvents,
 and temperature influence over the intermolecular forces measured by the $AFM$.  On the theoretical side,
we review a recent advancement made by the author in the coming
fields of quantum chemistry at finite temperature. Starting from
$\it{Bloch}$ equation, we have derived the sets of
 hierarchy equations for the reduced density operators in both canonical and grand canonical ensembles.
 They provide a law according to which the reduced density operators vary in temperature for the identical
 and interacting many-body particles. By taking the independent particle approximation, we have solved
 the equation in the case of a grand canonical ensemble, and obtained an eigenequation for the molecular
 orbitals  at finite temperature. The explicit
 expression for the temperature-dependent $\it{Fock}$ operator is also given. They will form a
 foundation for the study of the molecular electronic structures and their interplay with the finite temperature.
Furthermore, we clarify the physics concerning the temperature
effect on the electronic structure or processes of the molecules
which is crucial for both theoretical understanding and
computational study. Finally, we summarize our discussion and
point out the theoretical and computational issues for the future
explorations in the fields of quantum chemistry at finite
temperature.
\end{abstract}


\vspace{0.35in}
\section{Introduction}

 The history for quantum chemistry development is almost synchronous to that of quantum mechanics itself.
 It begins with $\it{Heitler}$ and $\it{London}$'s study of electronic structure of $H_{2}$
 molecule shortly after the establishment of wave mechanics for quantum particles~\cite{heitler}.
  There are two major types of molecular electronic theories: valence bond approach $\it{vs.}$ molecular orbital method
 with the latter being the popular one for the present investigation. It has gone through the stages from the evaluation
 of molecular integrals via a semiempirical way to the one by an $\it{ab\ initio}$ method. Correlation issue is always
 a bottleneck for the computational quantum chemistry and is under intensive study for over fifty years. For large
 molecular systems such as biomolecules and molecular materials, the development of the combined $\it{QM/MM}$ approach,
 pseudopotential method and linear scaling algorithm has significantly advanced our understanding of
  their structure and dynamics. There are about eight Nobel prize laureates whose researches are related to the molecular
  electronic structure theory. This not only recognizes the most eminent scientists who have made the outstanding
  contributions to the fields of quantum chemistry, but more importantly, it indicates the essential roles
 the electronic structure theory has been playing in the theoretical
chemistry as well as for the whole areas of molecular sciences.
 Nowadays, quantum chemistry has been becoming a maturing science~\cite{wei1,wei2}.

 Nevertheless, the current fields of quantum chemistry are only
 part of the story for the molecular electronic structure theory. From the pedagogical points of view,
 the quantum mechanics based on which the traditional quantum chemistry is built is a special case of more
 general quantum statistical mechanics~\cite{wei3,wei4,lowdin2}. In reality, the experimental observations are performed
 under the conditions with thermodynamic constraints. Henceforth, there is a need to extend the current areas of quantum
 chemistry to the realm of, for instance, finite temperature~\cite{wei3,wei4,lowdin2}.

  Indeed, many experimental investigations for various fields and for different systems have already shown the
  temperature or pressure effects on their microscopic structures [7-30,49-57,63,71-87]. The polymeric molecules are one
  of the most interesting systems for this sort of studies [7-16]. The
  experimental measurement on the absorption spectra,
  photoluminescence ($\it{PL}$), and photoluminescence excitation ($\it{PLE}$), and spectral line narrowing
  ($\it{SLN}$) for the $\it{PPV}$ and its derivatives all show the same trend of the blue shift with the increasing
  temperature~\cite{hagler,friend,yu1}. This attributes to the temperature dependence of their very rich intrinsic
  structures such as the vibronic coupling~\cite{yu2,deleuze1,deleuze2}. The experimental investigation of the
  temperature effect on the biomolecules started in the late nineteenth century~\cite{rakow,waugh}. Most recently, it
  has been extended to the study of folding and unfolding of protein or $\it{DNA}$~\cite{gaub0,williams,law}. In
  addition to the observed patterns for the unfolding forces with respect to the extension or temperature, it has been
  proved that the temperature-induced unfolding is another way for the study of mechanisms or pathways of protein
  folding or unfolding processes [19-23]. The newest related development is on the $\it{AFM}$ measurement made by
  $\it{Lo}$ et al. of the intermolecular forces for the biotin-avidin system in the temperature range from $286$ to
  $310 K$~\cite{simons}. It has shown that an increase of temperature will almost linearly decrease the strength of the
  bond rupture force for the individual biotin-avidin pair. The study of temperature effect on the absorption spectra
  of solvated electron began in the 1950's and it is still of current interest. A striking effect is that an increasing
  temperature will cause the positions of their maximal absorption red shift [71-85].

 In recent papers, we have deduced an eigenequation for the molecular orbitals~\cite{wei3,wei4}. It is the extension
 from the usual $\it{Hartree-Fock}$ equation at zero temperature to the one at any finite temperature~\cite{hartree,fock}.
 It opens an avenue for the study of the temperature effects on the electronic structures as well as their interplay with
  the thermodynamic properties. In the third section, we will present this equation and give the details for its
  derivation. In the next section, we will show four major types of experiments related to the study of the temperature
  influences over the microscopic structure of molecular systems. In the final section, we will discuss and analyze our
  presentations, and point out both theoretical and computational issues for the future investigation.

 \vspace{0.35in}
\section{Experimental Development}
 In this section, we mainly describe the experimental investigations related to the temperature effect
 on the bonding, structure and electronic spectra of molecules. We choose four kinds of the most
 recent developments in these fields which are of chemical or biomolecular interests.

 \subsection{Temperature effects on geometric structure and UV-visible electronic spectra of polymers}

 The first important systems where the important issues related to the temperature effect on the geometric structure
 and electronic spectra are the polymeric molecules. Many experimental investigations and some theoretical work already
 exist in the literature [7-16]. However, how the temperature changes the microscopic structures of the polymers
 are still not completely understood and there are many unresolved issues in interpreting their electronic spectra.
 We list here a few very interesting experimental investigations for the purpose of demonstration.

 The poly($\it{p}$-phenylenevinylene)($\it{PPV}$) is one of the prototype polymeric systems for the study of their
 various mechanical, electronic, and optical properties. The impact from the temperature on the
 absorption spectra, photoluminescence ($\it{PL}$), and photoluminescence excitation ($\it{PLE}$) of the $\it{PPV}$ have
 also been investigated both experimentally and theoretically~\cite{hagler,friend,yu1}. In an experiment carried out by
 $\it{Yu}$ et al., the absorption spectra are measured for the $\it{PPV}$ sample from the temperature $10$ to $330 K$. The
 details of the experiment are given in their paper~\cite{yu1}. The resulting spectra for the
 absorption at $T = 80$ and $300 K$ are shown in Figure 1 of that paper. We see that there
 is a pronounced change in the spectra when increasing the temperature. They also study the $\it{PL}$ and $\it{PLE}$
 spectra for the $\it{PPV}$. The measured $\it{PL}$ spectra at two temperatures: $77$ and $300 K$ are demonstrated
 in the Figure 3, and the $\it{PLE}$ spectra at those temperatures are depicted in the Figure 4 of the paper~\cite{yu1}.
 They both show the dramatic changes of the band blue shift when the temperature is increased. Similar studies have also
 been performed before by the other groups~\cite{hagler,friend}. They observed the similar behaviors.

Another interesting investigation is related to the temperature
effect on the spectral line narrowing ($\it{SLN}$) of the
poly(2-methoxy-5-($2^{'}$-ethylhexyloxy)-1,4-phenylenevinylene)($\it{MEH-PPV}$)
spin-coated from either $\it{THF}$ or $\it{CB}$
solvents~\cite{sheridan}. In the experiment done by Sheridan et
al., the $\it{SLN}$ is measured together with the absorption and
$\it{PL}$ as shown in Fig. 1 of their paper. It is found that the
same trend of the $\it{SLN}$ blue shift is observed as that for
the absorption and $\it{PL}$ with an increasing temperature. They
attribute this to the same reason of the electronic structure
modification resulting from the variation of the temperature.

 \subsection{Temperature effects on structure, dynamics, and
 folding/unfolding of biomolecules}
  Biomolecules are complex systems, featuring a large molecular size, a heterogeneity of atomic constitutes and
 a variety of conformations or configurations. Their energy landscape thereby exhibits multiple substates and multiple
energy barrier, and varies in size for the barrier
heights~\cite{wolynes1,shakhnovich,karplus1,dill,thirumalai}.
 The temperature should have a strong effect on their structure and dynamics including the folding or
 unfolding [17-57]. This effect could be either from the fluctuation of thermal motion of molecules or due to
 the redistribution of electronic charge as we will discuss in the next section.

 The experimental observation of the temperature impact on the
microscopic structure of biological systems dates back to the very
early days. One focus, for example, is on the measurement of the
elastic properties of the human red blood cell membrane as a
function of temperature~\cite{rakow,waugh}. Another related study
is about the influence over the thermal structural transition of
the young or unfractionated red blood cells due to the involvement
of protein spectrin which might modified the spectrin-membrane
interaction~\cite{minetti1,minetti2}. Most recently, the atomic
force microscopy ($\it{AFM}$) has been used to detect the impact
from the variation of the temperature on the spectrin protein
unfolding force as well as on the bond rupture force for the
biotin-avidin system~\cite{gaub0,williams,law,simons}.

 The $\it{AFM}$ is a surface imaging technique with an atomic-scale resolution
   capable of measuring $\it{any}$ types of the forces as small as $10^{-18}\ N$. It combines the principle of scanning
  tunnelling microscopes ($STM$) and stylus profilometer, and therefore can probe the surfaces of both conducting and
  nonconducting samples~\cite{binnig1,binnig2}. The imaging on soft materials such as biomolecules with the $\it{AFM}$
  has been performed beginning in the 1980's~\cite{hansma1,hansma2,gaub1}. Recently, it has been applied to measure
  the adhesive forces and energies between the biotin and avidin pair also as we will show
in the next subsection~\cite{moy1,moy2,moy3,beebe1}. Unlike other
experimental techniques, the $\it{AFM}$ features a high precision
  and sensitivity to probe the surface with a molecular resolution, and can be done in physiological environments.

 In an $\it{AFM}$ investigation of mechanical unfolding of titin protein, for example, the restoring forces all show a
 sawtooth like pattern with a definite periodicity. It reveals much information about
 the mechanism of unfolding processes~\cite{rief,fernandez1}. The
 observed pattern, in addition to fit a worm-like chain model, has also been verified by the steered molecular
 dynamics or Monte carlo simulations~\cite{schulten1,discher2}. Similar study has been
 extended to other systems~\cite{fernandez2,lenne,discher3}.

 The same kind of experiments has also been performed by varying the temperature. In the experiment carried out by Spider
  and Discher et al.~\cite{law}, the spectrin protein is chosen for the $\it{AFM}$ study at different temperatures.
 Thousands of tip-to-surface contacts are performed for a given temperature because of the statistical nature of the
 $\it{AFM}$ measurement. The observed curve for the relation between the unfolding force and extension shows the
 similar sawtooth pattern for all temperatures. In addition, the
  tandem repeat unfolding events are more favored at lower temperature as demonstrated in the unfolding length histograms. Most striking is that the unfolding
  forces show a dramatically nonlinear decreasing relation as the temperature $T$ approaches the transition
 temperature $T_{m}$. This is shown in the Figure $3B$ of paper~\cite{law}.

 Similar behaviors regarding the force-temperature dependence have also been observed via either $\it{AFM}$ or
  optical tweezers for the forced overstretching transition for the individual double-stranded $\it{DNA}$
  molecules~\cite{gaub0,williams,bloomfield1,bloomfield2}.

 Some other interesting experiments which illustrate the effect of temperature on the microscopic structure of
 biomolecules have also been performed even though the detailed physical origins of the effect (from either the electrons
 or the molecules) have not been specified~\cite{eriksson,huang,wang1,wang2,mayer,dyer}. In a circular dichroism
 ($\it{CD}$) spectra and high resolution $\it{NMR}$ study, for instance, it shows that the secondary structure of the
 Alzheimer $\beta$ (12-28) peptide is temperature-dependent with an extended left-handed
 $3_{1}$ helix interconverting with a flexible random coil conformation~\cite{eriksson}. Another example
 is related to the analysis of the temperature-dependent interaction of the protein $\it{Ssh10b}$ with $\it{DNA}$ which
 influences the $\it{DNA}$ topology~\cite{huang,wang1,wang2}. The study from the heteronuclear $\it{NMR}$ and
 site-directed mutagenesis indicates that the $\it{Ssh10b}$ exists as a dimer: $T$ form and $C$ form. Their ratio is
 determined by the $Leu^{61}-Pro^{62}$ peptide bond of the $\it{Ssh10b}$ which is sensitive to the temperature.

 \subsection{Temperature effects on intermolecular forces}
 The study of the general issues related to the temperature effects on the microscopic structure has been most
 recently extended to the realm of intermolecular forces. Since the usual intermolecular forces
  such as hydrogen bond, $\it{van\ der\ Waals}$ force, ionic bond, and hydrophobic interaction are weak and typically
  of the order of $\it{0.1\ eV}$ or $\it{4.0\ kT}$ at the physiological temperature,
  the variation of temperature will thereby have a very strong influence over the strength of these forces.

 The first experimental investigation on the temperature-dependent intermolecular forces is for the
 biotin-avidin system and by an $AFM$ measurement~\cite{simons}. The biotin-avidin complex is a prototypical receptor
 and ligand system with the biotin binding strongly up to four avidin
 protein~\cite{green,darst,sussman,bolognesi1,bolognesi2}. They have an extremely high binding affinity, and therefore
 serves as a model system for various experimental investigations.
 In the experiment carried out in $\it{Beebe}$'s group~\cite{simons}, the receptor avidin
 is attached to the  $\it{AFM}$ tip and linked to the agarose bead functionalized with the biotin.
  The temperature of the entire $\it{AFM}$ apparatus is varied at a range from
  $286$ to $310 K$. In addition, the loading rate is kept very slow so that the thermal equilibrium for the
  biotin-avidin pairs is assumed. The forces expected to be determined is the rupture force $F_{i}$ between
  the individual biotin-avidin pair which is defined as the maximum restoring force~\cite{simons}. In actual $AFM$
  experiment, however, the total adhesive force between the tip and substrate is measured. It is a
  sum of finite number $n$ of the interactions between each biotin and avidin pair. To extract the individual and
  average bond rupture, a statistical method has been developed in $\it{Beebe}$'s
 group~\cite{beebe1,beebe2,beebe3,beebe4,bai1}. They assume a $\it{Poisson}$ distribution for the number $n$ of the
 discrete rupture forces or linkages from multiple measurements, and have obtained the single force $F_{i}$ at
  different temperatures. The result is shown in the Figure $3$ of the paper~\cite{simons}. We see that the individual
  rupture force $F_{i}$ for the biotin-avidin pair is decreased by about five-fold in strength when the temperature is
  increased from $286$ to $310 K$.

  To interpret the observed temperature impact on the biotin-avidin forces, $\it{Peebe}$'s group has performed a
 thermodynamic analysis~\cite{simons}. Based on the simple models and arguments~\cite{tavan,schulten}, they have
 come out an equation that connects the square of the single bond-rupture force $F_{i}$ to the absolute temperature
 $T$ as follows,
 \begin{equation}
  F^{2}_{i} = 2\Delta E^{\ddagger} k_{bond} - 2k_{B} T k_{bond}
  \ln\left(\frac{\tau_{R}}{\tau_{D}}\right)
\end{equation}
 where the $k_{bond}$ is the force constant of the individual biotin-avidin pair, and the time $\tau_{R}$ is the
 characteristic time needed to break $n$ pairs of those forces. The $E^{\ddagger}$ is the energy required to remove
 the biotin from avidin's strongest binding site and the corresponding time is $\tau_{D}$.  More detailed on this
 analysis can be found in the paper~\cite{simons}. The relation between the square of the force $F_{i}$ and temperature
 $T$ is also plotted as Figure $5$ in that paper. Therefore, from the relation (1) and that Figure, the information
  about the stiffness of the ligand and receptor bond and the critical binding energy, etc. can be obtained.
 Obviously, what we need at present is a microscopic theory which can account for all these relations and properties.

\subsection{Temperature effects on absorption spectra of electrons in solvents}

  The structure and dynamics of solute in solvent is one of the most important fields in chemistry since most of the
  chemical reactions occur in solution phases. In the meantime, it is also one of the most challenging fields in
   theoretical chemistry with many unsettled issues. The variation of temperature in the measurement of absorption
  spectra of solvated electron in various solvents has proved to be a useful means for the understanding of the solvation
     processes [71-85].

 There are several experimental techniques available for this type of studies with the pulse radiolysis being the most
 commonly used one. There are also several research groups doing the similar experimental investigations and obtaining
 the consistent results relating to the temperature effects on the optical absorption spectra of solvated electron in
 solvents. In a recent experiment carried out in $\it{Katsumura}$'s group, for example, the pulse radiolysis technique is
 employed to study the optical absorption spectra of the solvated electron in the ethylene glycol at different
 temperatures from $290$ to $598 K$ at a fixed pressure of $100$ atm. In addition to the faster
decay of absorptions, it is found that, their maximal positions
shift to the red with the increasing temperature as shown in the
 Figures 1 to 3 of the paper~\cite{katsumura4}.
 This is in contrast to the situation for the electronic spectra of the polymers. They also point out the need to
 quantify the change of the density in the experiment in order to really understand the observed results.

 The same type of experiment has been extended to the study of the optical absorption spectra for $Ag^{0}$ and
  $Ag_{2}^{+}$ in water by varying the temperature, and similar results have been obtained~\cite{katsumura3}.

 \vspace{0.35in}
\section{Theoretical Development}
 Having presented four different types of experiments above, we have seen that the temperature effect on the microscopic
 structure of molecules is a very interesting and sophisticated field. Many more needs to be understood and probed.
 Even though the experimental study has been for a long time, very limited number of the related theoretical work
 is available, especially at the first-principle level. In other words, the quantum chemistry at finite temperature
 is not a well-defined or well-established field~\cite{wei3,wei4,lowdin2}.

It is true that the influence of temperature on the microscopic
structure is a complicated phenomenon. There exists different
functioning mechanisms. One consideration is that the variation of
the temperature, according to the Fermi-Dirac statistics, will
change the thermal probability distribution of single-particle
states for a free electron gas. It is expected that similar
situation should occur for an interacting electron system, and
therefore its microscopic structure will be correspondingly
altered. Another consideration is that, for molecules or solids,
the thermal excitation will cause the change of the time scales
for the molecular motions. This will most likely bring about the
transitions of the electronic states, and therefore lead to the
breakdown of the Born-Oppenheimer approximation. Electron-phonon
interaction is a fundamental topic in solid state physics and its
temperature dependence is well-known. As a result, the temperature
could change the strength of the coupling between the electronic
and molecular motions. Nevertheless, we tackle the issues in a
simpler way. We treat only an interacting identical $\it{fermion}$
system, or neglect the coupling of the electronic motion with
those of the nucleus in molecules or solids. We expect that some
sort of the general conclusions will come out from this study. As
a matter of fact, this is also the approach usually adopted in a
non-adiabatic molecular dynamics, in which purely solving the
eigenequation for the electrons will provide the reference states
for the investigation of the coupling motions between the
electrons and the nucleus of the molecules.

 In the following, we will present a self-consistent equation within the framework of equilibrium
 statistical mechanics which decides the molecular orbitals at a given temperature.

 \subsection{Hierarchy Bloch equations for reduced density
 operators in canonical ensemble}
 We consider an identical and interacting $N$-particle system. In a canonical ensemble, its
 $N$th-order density operator takes the form
 \begin{equation}
   D^{N} = \exp(-\beta H_{N}),
\end{equation}
 and satisfies the $\it{Bloch}$ equation~\cite{bloch,kirkwood}
 \begin{equation}
-\frac{\partial}{\partial\beta} D^{N} = H_{N} D^{N},
\end{equation}
 where
 \begin{equation}
  H_{N} = \sum_{i=1}^{N} h(i) + \sum_{i<j}^{N} g(i,j),
\end{equation}
 is the $\it{Hamiltonian}$ for the $N$ particle system composed of the one-particle operator $h$ and two-body
 operator $g$. The $\beta$ is the inverse of the product of $\it{Boltzmann}$ constant $k_{B}$ and absolute
  temperature $T$.

 Since the $\it{Hamiltonian}$ (4) can be written as a reduced $\it{two}$-body operator form,
 the second-order reduced density operator suffices to describe its $N\ (\ge 2)$ particle quantum states. A $p$th-order
 reduced density operator is generally defined
 by~\cite{kummer,harriman1}
 \begin{equation}
D^{p} = L^{p}_{N} (D^{N}),
\end{equation}
 where $L^{p}_{N}$ is the contraction operator acting on an $N$th-order tensor
 in the $N$-particle $\it{Hilbert}$ space $V^{N}$. The trace of the $D^{p}$ gives the partition
 function,
\begin{equation}
 Tr (D^{p}) = Z (\beta, V, N).
\end{equation}
  Rewrite the $\it{Hamiltonian}$ in a form
 \begin{equation}
H_{N}=H^{p}_{1}+\sum_{j=p+1}^{N}h(i)+\sum_{i=1}^{p}\sum_{j=p+1}^{N}g(i,j)
+\sum_{i<j\\(i\ge p+1)}^{N} g(i,j),
\end{equation}
where
\begin{equation}
  H_{1}^{p} = \sum_{i=1}^{p} h(i) + \sum_{i<j}^{p} g(i,j),
\end{equation}
 and apply the contraction operator $L^{p}_{N}$ on both sides of the Eq. (3), we develop
 an equation that the $p$th-order density operator
 satisfies~\cite{wei3}
 \begin{eqnarray}
\nonumber
-\frac{\partial}{\partial\beta}D^{p}&=&H^{p}_{1}D^{p}+(N-p)L^{p}_{p+1}
\left[h(p+1)D^{p+1}\right]+(N-p)L_{p+1}^{p}\left[\sum_{i=1}^{p}g(i,p+1)D^{p+1}\right]+\\
&&+\left(\begin{array}{c} N-p\\2\end{array}\right)
L_{p+2}^{p}\left[g(p+1,p+2)D^{p+2}\right].
\end{eqnarray}
 It provides a law according to which the reduced density operators vary
 in terms of the change of temperature.

 \subsection{Hierarchy Bloch equations for reduced density
 operators in grand canonical ensemble}
 The above scheme for deducing the equations for the reduced operators can be readily extended to
  the case of a grand canonical ensemble, which is a more general one with a
 fluctuating particle number $N$. In this ensemble, the density operator is defined in the entire
  $\it{Fock}$ space
 \[ F = \sum_{N=0}^{\infty} \oplus V^{N},   \]
  and is written as the direct sum of the density operators $D_{G}(N)$
 associated with the $N$-particle $\it{Hilbert}$ space $V^{N}$,
\begin{equation}
   D_{G} = \sum_{N=0}^{\infty} \oplus D_{G}(N),
\end{equation}
 where
 \begin{eqnarray}
\nonumber D_{G}(N)&=& \exp[-\beta(H-\mu N)],\\
  &=& \exp(-\beta \bar{H}),
\end{eqnarray}
and
\begin{equation}
  \bar{H} = H - \mu N,
\end{equation}
is called the grand $\it{Hamiltonian}$ on $V^{N}$. The form of the
$\it{Hamiltonian}$ $H$ has been given by Eq. (4) and the $\mu$ is
the chemical potential. The corresponding $p$th-order reduced
density operator is therefore defined as
\begin{equation}
 D^{p}_{G} =\sum_{N=p}^{\infty}\oplus
 \left(\begin{array}{c}N\\p\end{array}\right)L_{N}^{p}[D_{G}(N)],
\end{equation}
with the trace given by
\begin{equation}
  Tr(D_{G}^{p}) =
  \left<\left(\begin{array}{c}N\\p\end{array}\right)\right>
  D^{0}_{G}
\end{equation}
 and
\begin{equation}
  D_{G}^{0} = \Xi(\beta, \mu, V).
\end{equation}
 The $\Xi(\beta, \mu, V)$ is the grand partition function.

 In a similar manner, we can also derive the hierarchy equations that
 the reduced density operators in the grand canonical ensemble
 obey~\cite{wei4}
\begin{eqnarray}
\nonumber
-\frac{\partial}{\partial\beta}D^{p}&=&\bar{H}^{p}_{1}D^{p}+(p+1)L^{p}_{p+1}
\left[\bar{h}(p+1)D^{p+1}\right]+(p+1)L_{p+1}^{p}\left[\sum_{i=1}^{p}g(i,p+1)D^{p+1}\right]+\\
&&+\left(\begin{array}{c} p+2\\2\end{array}\right)
L_{p+2}^{p}\left[g(p+1,p+2)D^{p+2}\right],
\end{eqnarray}
where
\begin{equation}
  \bar{H}_{1}^{p} = \sum_{i=1}^{p}\bar{h}(i) + \sum_{i<j}^{p}
  g(i,j),
\end{equation}
and
\begin{equation}
 \bar{h}(i) = h(i)-\mu.
\end{equation}
It gives us a law with which the reduced density operators in the
grand canonical ensemble vary in temperature.

 \subsection{Orbital approximation and Hartree-Fock equation at
  finite temperature}
  The Eqs. (9) and (16) define a set of $\it{hierarchy}$ equations that establish the relation among the reduced density
 operators $D^{p}$, $D^{p+1}$, and $D^{p+2}$. They can be solved either in an exact scheme or by an approximate
 method. The previous study of $N$ electrons with an independent particle approximation to the
 $\it{Schr}$$\ddot{o}$$\it{dinger}$ equation for their $\emph{pure}$ states has lead to the $\it{Hartree-Fock}$ equation
 for the molecular orbitals [94-100]. We thereby expect that the same approximate scheme to the reduced $\it{Bloch}$
 equations (9) or (16), which hold for more general mixed states, will yield more generic eigenequations than the usual
  $\it{Hartree-Fock}$ equation for the molecular orbitals.

 We consider the case of a grand canonical ensemble. When $p = 1$, Eq. (16) reads
\begin{equation}
-\frac{\partial}{\partial\beta}
D^{1}=\bar{H}_{1}D^{1}+\frac{Tr(\bar{h}D^{1})}{D^{0}}D^{1}-\frac{1}{D^{0}}D^{1}\bar{h}
D^{1}+2L^{1}_{2}\left[g(1,2)D^{2}\right]+3L^{1}_{3}\left[g(2,3)D^{3}\right].
\end{equation}
  Under the orbital approximation, the above second-order and third-order reduced density
  operators for the electrons can be written as
  \begin{equation}
\nonumber D^{3} = D^{1}\wedge D^{1}\wedge D^{1}/(D^{0})^{2}
 \end{equation}
  and
 \begin{equation}
 D^{2} = D^{1}\wedge D^{1}/D^{0}.
\end{equation}
These are the special situations for the statement that a
$p$th-order reduced density matrix can be expressed as a $p$-fold
 $\it{Grassmann}$ product of its first-order reduced density matrices.
 With these approximations, the last two terms of Eq. (19) can be evaluated in a straightforward way as
follows
\begin{equation}
 2L^{1}_{2}\left[g(1,2)D^{2}\right] = (J-K)D^{1},
\end{equation}
and
\begin{equation}
 3L^{1}_{3}\left[g(2,3)D^{3}\right]
 =\frac{Tr(gD^{2})}{D^{0}}-\frac{1}{D^{0}}D^{1}(J-K)D^{1},
\end{equation}
where
\begin{equation}
J = Tr_{2}\left[g\cdot D^{1}(2;2)\right]/D^{0},
\end{equation}
and
\begin{equation}
 K = Tr_{2}\left[g\cdot (2,3)\cdot D^{1}(2;2)\right]/D^{0},
\end{equation}
are called the $\it{Coulomb}$ and exchange operators,
respectively. With the $(2,3)$ being the exchange between the
particle $2$ and $3$, the action of the $K$ on the reduced density
operator is
\begin{eqnarray}
\nonumber K\cdot D^{1}(3;3)&=& Tr_{2}\left[g\cdot (2,3)\cdot
D^{1}(2;2)\right]
/D^{0}\cdot D^{1}(3;3) \\
&=& Tr_{2}\left[g\cdot D^{1}(3;2)\cdot D^{1}(2;3)\right]/D^{0}.
\end{eqnarray}
Substitution of Eqs. (22) and (23) into Eq. (19) yields the
$\it{Bloch}$ equation for the first-order reduced density matrix
of $N$ interacting electrons under orbital approximation,
\begin{equation}
-\frac{\partial}{\partial\beta}D^{1}=(F-\mu)D^{1}+\left(\frac{Tr\bar{h}D^{1}}{D^{0}}
+\frac{TrD^{2}}{D^{0}}\right)D^{1}-\frac{1}{D^{0}}D^{1}(F-\mu)D^{1},
\end{equation}
where
\begin{equation}
    F = h+J-K,
\end{equation}
is called the $\it{Fock}$ operator at finite temperature. Redefine
the normalized first-order reduced density operator
\begin{equation}
 \rho^{1} = D^{1}/D^{0},
\end{equation}
we can simplify above equation into
\begin{equation}
-\frac{\partial}{\partial\beta} \rho^{1}
=(F-\mu)\rho^{1}-\rho^{1}(F-\mu)\rho^{1}.
\end{equation}

Furthermore, from Eq. (30) and its conjugate, we get
\begin{equation}
  F\rho^{1}-\rho^{1}F = 0,
\end{equation}
 which means that the $\it{Fock}$ operator $F$ and the first-order reduced density matrix $\rho^{1}$ commute.
 They are also $\it{Hermitian}$, and therefore have common eigenvectors $\{|\phi_{i}\rangle\}$. These vectors are
 determined by the following eigen equation for the $\it{Fock}$ operator,
\begin{equation}
  F|\phi_{i}\rangle = \epsilon_{i}|\phi_{i}\rangle.
\end{equation}

The first-order reduced density operator is correspondingly
expressed as
\begin{equation}
  \rho^{1} =
  \sum_{i}\omega(\beta,\mu,\epsilon_{i})|\phi_{i}\rangle\langle\phi_{i}|,
\end{equation}
where $\omega(\beta,\mu,\epsilon_{i})$ is the thermal probability
that the orbital is found to be in the state
$\{|\phi_{i}\rangle\}$ at finite temperature $T$. Substituting Eq.
(33) into Eq. (30), we can obtain the equation this thermal
probability $\omega(\beta,\mu,\epsilon_{i})$ satisfies,
\begin{equation}
-\frac{\partial}{\partial\beta} \omega(\beta, \mu, \epsilon_{i})=
(\epsilon_{i}-\mu)\omega(\beta,\mu,\epsilon_{i})-(\epsilon_{i}-\mu)\omega^{2}
(\beta,\mu,\epsilon_{i}).
\end{equation}
Its solution has the same usual form of the $\it{Fermi-Dirac}$
statistics for the free electron gas as follows,
\begin{equation}
\omega(\beta,\mu,\epsilon_{i})=\frac{1}{1+e^{\beta(\epsilon_{i}-\mu)}},
\end{equation}
with the energy levels $\{\epsilon_{i}\}$ determined by Eq. (32).

\vspace{0.35in}
\section{Discussion, Summary and Outlook}
In this paper, we have presented a description of both
experimental and theoretical developments related to the
temperature impact on the microscopic structure and processes for
the molecules.

In the theoretical part of this paper, we have depicted the sets
of hierarchy $\it{Bloch}$ equations for the reduced statistical
density operators in both a canonical and a grand canonical
ensembles for the identical fermion system with two-body
interaction. We have solved the equations in the latter case under
a single-orbital approximation and obtained an eigen-equation for
the single-particle states. It is the extension of usual commonly
used $\it{Hartree-Fock}$ equation at the absolute zero temperature
to the situation at any finite temperature. The average occupation
number formula for each single-particle state is also obtained,
which has the same analytical form as that for the free electron
gas with the single-particle state energy determined by the
$\it{Hartree-Fock}$ equation at finite temperature (32).

From Eqs. (24), (25) and (28), we see that the $\it{Coulomb}$
operator $J$, the exchange operator $K$, and therefore the
$\it{Fock}$ operator $F$ are both the coherent and the incoherent
superpositions of single-particle states. They are all
temperature-dependent through an incoherent superposition factor,
the $\it{Fermi-Dirac}$ distribution,
$\omega(\beta,\mu,\epsilon_{i})$. Therefore, the mean force or the
force field, and the corresponding microscopic structure are
temperature-dependent.

 We have expounded the physics relating to the temperature effects
 on the electronic structure or processes of the molecules. This
 is very critical for our understanding and studying the
 temperature influence over the molecular structure. From this
 analysis, for example, we could see that the temperature should have a
 stronger effect on the molecular transition states and therefore their chemical reactivity.
 Starting from Eqs. (9) or (16), it will be a very significant work to establish a
 corresponding multireference theory for the molecular orbitals at finite temperature~\cite{schlosser,wahl,gilbert}.

  On the experimental sides, we have exposed four major fields of investigations of chemical or biomolecular interests,
  which show the temperature impact on their structures, spectra, or bond rupture forces.

  The complete determination of the geometric structure and electronic spectra of the polymeric molecules is a very
  difficult task. As has been stated in papers~\cite{deleuze1,deleuze2}, there are many different components
  contributing to the change of the spectra. At present, we focus on the study of the effect from the temperature.
  We have demonstrated that it can alter both the shapes and positions of the absorption and other spectra for
  the $\it{PPV}$ and its derivatives. As has been analyzed, the increase of the temperature will cause the excitation of
  the vibrational, rotational and liberal motions which might also lead to the electronic transition. The
  $\it{Huang-Rhys}$ parameter has been introduced to describe the strength of the coupling between the electronic
  ground- and excited-state geometries. Furthermore, it has been found that this factor is an increasing function of
  the temperature~\cite{hagler,friend,yu1}. Obviously, a more detailed study of the electronic structure,
 excitation and spectroscopic signature at the first-principle level which includes the temperature-dependent
force field is expected.

The temperature has proved to be a big player in both experimental
and theoretical study of the structure and dynamics of
biomolecules including their folding or unfolding. At a first
glance, the energy gap between the $\it{HOMO}$ and $\it{LUMO}$ for
the biomolecules should be small or comparable to the
$\it{Boltzmann}$ thermal energy $k_{B}T$ because of
  their very large molecular size. Therefore a change of the temperature should have a strong influence over their
  electronic states, and consequently, the energy landscape and the related dynamics including the folding or unfolding,
  etc. The experimental investigation with the $\it{AFM}$ and other techniques of the temperature effect on the shift of
  their unfolding pathways might have verified this sort of thermal deformation of potential energy
  landscape~\cite{gaub0,williams,law}. This is in contrast to the tilt and deformation  of energy landscape of
  biomolecules including their transition states resulting from the applied mechanical forces~\cite{evans1}.

  Unfolding proteins by temperature is not just one of the classical experimental techniques
 for the study of the structure, dynamics and energetics of the biomolecules. It has also been employed, for example,
 in the molecular dynamic
 simulation to study the structure of transition states of $\it{CI2}$ in water at two different temperatures: $298\ K$
 and $498\ K$~\cite{daggett}. The later high temperature is required in order to destabilize the
 native state for monitoring the unfolding as done in the real
 experiments. In another recent molecular dynamics simulation~\cite{karplus2}, $\it{Karplus}$'s
 group has compared the temperature-induced unfolding with the
 force-stretching unfolding for two $\beta$-sandwich proteins and two $\alpha$-helical proteins.
  They have found that there are the significant differences in the unfolding pathways for two approaches.
  Nevertheless, in order to get more reliable results, temperature-dependent force fields need to be developed and
  included in the molecular dynamics simulations. This is also the case in the theoretical investigation of protein
  folding since an accurate simulation of protein folding pathways requires better stochastic or temperature-dependent
  potentials which have become the bottleneck in structure
  prediction~\cite{wolynes1,shakhnovich,karplus1,dill,thirumalai}. From structural points of view, the variation of
  temperature leads to the change of the mean force or the energy landscape, and therefore provides a vast variety of
  possibilities, for instance, in the protein design and engineering.

 The intermolecular forces are ubiquitous in nature. They are also extremely important for
 the biological systems and for the existence of life. The intermolecular forces have the specificity which are
 responsible for the molecular recognition between receptor and ligand, antibody and antigen, and complementary
 strands of $\it{DNA}$, and therefore for the regulation of complex organization of life~\cite{frauenfelder}.
 For these reasons, the experiment carried out in $\it{Beebe}$'s group has an immediate significance. It has demonstrated
 that the temperature can be an important factor for changing the specificity of the intermolecular forces and
 therefore the function of life~\cite{simons}. Nevertheless, how the charge redistribution occurs due to the variation
 of the temperature has not been interpreted, and a microscopic theory for quantifying the temperature influence on the
 intermolecular forces is still lacking. Since the delicate study of the intermolecular forces provides the insight into
 complex mechanisms of ligand-receptor binding and unbinding processes or pathways,
 a paramount future research is to establish the links between the
  intermolecular forces and the temperature within the quantum many-body theory.

 The theoretical study of the temperature effects on the optical absorption spectra of solvated electron is still in very
 early stage and few published works are available~\cite{jortner,brodsky,berne,nicolas}. One of the earliest
 studies by $\it{Jortner}$ used a cavity model to simulate the solvated electron where the electron is confined to
 the cavity surrounded by the dielectric continuum solvent~\cite{jortner}. However, his study is not of fully
 microscopic in nature since he assumed a temperature dependence of phenomenological dielectric constants
 which were also obtained from the available experimental data. In addition, the model used is too simplified
 and, for instance, it neglects the intrinsic structure of the solvent molecules. There are a few recent
 investigations on the temperature effects on the absorption spectra of solvated electrons. They all cannot catch the
 full features of the experimental observations. One reason is that the physical nature for the process is not totally
 understood which might leads to incorrect models used for the simulation. The other is to utilize the crude models
 which might have omitted some important physical effects. For example, in an analysis by $\it{Brodsky}$ and
 $\it{Tsarevsky}$~\cite{brodsky}, they have concluded a temperature-dependence relation for the
 spectra which is, however, in contradiction with the experimental findings at high temperature. The quantum path-integral
 molecular dynamics simulation cannot produce those temperature-dependence relations observed in the
 experiments~\cite{berne}. In a recent quantum-classical molecular-dynamics study by $\it{Nicolas\ et\ al}$, even though
 the temperature-dependent features of optical absorption spectra for the solvated electron in water have been
 recovered~\cite{nicolas}, however, they claim that the red shifts of absorption spectra with increasing temperature
 observed in both experiments and calculation are due to the density effect instead of temperature. This might cast the
 doubt of usefulness of our current theoretical work in this area. However, after examining their work, we observe that
 they actually have $\it{not}$ included any temperature effect on the electron in their theoretical model. This effect
 might be either from the $\it{Fermi-Dirac}$ distribution for individual electrons or due to the electronic excitation
 caused by the thermal excitation of the solvent, as we have discussed in paper~\cite{wei3}. Obviously, much finer
 theoretical work or more experimental investigations in this area are expected to resolve this dispute.

 In addition to the systems discussed above, there are other types
 which also show the temperature impact on their microscopic
structures. Either theoretical or experimental work have been done
or are in progress. Examples include the study of the temperature
dependence of the $\it{Coulumb}$ gap and the density of states for
the $\it{Coulomb}$ glass, the experimental investigation of the
temperature effects on the band-edge transition of $ZnCdBeSe$, and
the theoretical description of the influence from the temperature
on the polaron band narrowing in the oligo-acene crystals
~\cite{schreiber,hsieh,hannewald}.

  To sum up, the quantum chemistry at finite temperature is a new and exciting field. With the combination of the
  techniques from current quantum chemistry with those developed in statistical or solid state physics, it will provide
  us with a myriad number of opportunities for the exploration.




\end{document}